\documentclass[conference]{IEEEtran}
\IEEEoverridecommandlockouts
\usepackage{cite}
\usepackage{amsmath,amssymb,amsfonts}
\usepackage{algorithmic}
\usepackage{graphicx}
\usepackage{textcomp}
\usepackage{xcolor}
\def\BibTeX{{\rm B\kern-.05em{\sc i\kern-.025em b}\kern-.08em
    T\kern-.1667em\lower.7ex\hbox{E}\kern-.125emX}}
\begin{document}

\title{Identification and Optimization of Redundant Code Using Large Language Models}

\author{\IEEEauthorblockN{Shamse Tasnim Cynthia}
\IEEEauthorblockA{\textit{Department of Computer Science} \\
\textit{University of Saskatchewan}\\
Saskatoon, Canada \\
shamse.cynthia@usask.ca\\
}
}

\maketitle

\begin{abstract}
Redundant code is a persistent challenge in software development that makes systems harder to maintain, scale, and update. It adds unnecessary complexity, hinders bug fixes, and increases technical debt. Despite their impact, removing redundant code manually is risky and error-prone, often introducing new bugs or missing dependencies. While studies highlight the prevalence and negative impact of redundant code, little focus has been given to Artificial Intelligence (AI) system codebases and the common patterns that cause redundancy. Additionally, the reasons behind developers unintentionally introducing redundant code remain largely unexplored.
This research addresses these gaps by leveraging large language models (LLMs) to automatically detect and optimize redundant code in AI projects. Our research aims to identify recurring patterns of redundancy and analyze their underlying causes, such as outdated practices or insufficient awareness of best coding principles. Additionally, we plan to propose an LLM agent that will facilitate the detection and refactoring of redundancies on a large scale while preserving original functionality.
This work advances the application of AI in identifying and optimizing redundant code, ultimately helping developers maintain cleaner, more readable, and scalable codebases.
\end{abstract}


\section{Problem statement}
In software development, maintaining high code quality is essential for the scalability, maintainability, and reliability of software projects \cite{van2002java}. However, redundant code poses a significant challenge to read and maintain while offering little to no added functionality, impacting the maintainability and evolution of software \cite{charalampidou2018structural}. Moreover, it frustrates the senior developers and increases the workload for maintaining the code \cite{qiong2020optimization}, making bug fixing, feature updates, and future improvements become more time-consuming and error-prone \cite{sundelin2020hidden}. The risk is even more pronounced in test case methods, where long, scenario-based code can violate coding best practices, leading to exaggerated and redundant test logic similar to issues in source code \cite{alegroth2017towards}. Furthermore, removing the redundant code is often non-trivial, and manually deleting them can lead to missed dependencies and encourage new bugs \cite{shackleton2023dead}. Thus, automatically identifying and optimizing redundant code is necessary to prevent additional challenges for developers while implementing new features or performing bug fixes.

Previous studies have analyzed source code to identify redundant code in terms of dead code or unused code and investigated their harmful impact on overall software quality \cite{romano2018multi, malavolta2023javascript, shackleton2023dead, dandan2012rc, suzuki2017exploratory, carneiro2024investigating}. For instance, Simone et al. \cite{romano2018multi} found that dead code is harmful both in the maintenance and design phases of software development.  
Shackleton et al. \cite{shackleton2023dead} noted that in large Python codebases, software evolution often leads to unused code and data, reducing efficiency and compromising user privacy. 
Similarly, Dandan et al. \cite{dandan2012rc} found that redundant code complicates debugging and increases the risk of software bugs. Qiong et al. \cite{qiong2020optimization} and Malavolta et al. \cite{malavolta2023javascript} focused on JavaScript, using both static and dynamic analyses to detect and remove redundant code.
Suzuki et al. \cite{suzuki2017exploratory} found significant functional redundancy in code repositories, identifying 984 redundant method pairs across 41.17\% of the analyzed projects. Similarly, Eduardo et al. \cite{carneiro2024investigating} showed that redundant code issues persist during refactoring and negatively impact code quality.
 
While studies show that redundant code is common and harms software quality, they fail to identify the coding patterns that contribute most to redundancy. Additionally, many studies focus solely on specific types of redundant code, such as dead code, leaving other forms unexplored. For instance, dead code refers to programming that is never executed during a program's runtime and remains unused \cite{romano2018dead}. In contrast, redundant code involves multiple locations containing identical or nearly identical statements, which may have a broader impact on maintainability and code efficiency. This highlights the need for a more comprehensive investigation into all types of redundant code.
Moreover, the reasons why developers introduce redundant code have not yet been thoroughly investigated, leaving a significant gap in understanding the root causes of these inefficiencies.

In our work, we aim to improve the overall quality and maintainability of open-source AI projects by identifying and optimizing redundant code. By leveraging LLMs, our research aims to automatically identify common patterns that contribute to redundancy in source code. Since AI systems often manage large-scale data and perform complex computations \cite{do2020cost}, it is imperative to reduce the redundancy in these systems to enhance maintainability and efficiency. With LLMs being widely adopted for code-related tasks \cite{zhang2024attacks, berabi2024deepcode, koziolek2024llm}, we plan to harness their capabilities to optimize code by effectively reducing redundancy. Furthermore, we seek to uncover the reasons why developers unintentionally introduce these inefficiencies. Through the creation of a framework capable of analyzing and optimizing these redundancies, this research will contribute to enhancing code readability, understandability, and maintainability, ultimately supporting the sustainable development of AI systems.

\textbf{Research Question:}
We aim to answer the following research questions:

    \textbf{RQ1:} To what extent does the source code of AI systems contain redundant code, and what is the impact of the redundant code on the overall code quality?
    
    \textbf{RQ2:} What are the common coding patterns that contribute to the introduction of redundant code in AI systems? 
    
    \textbf{RQ3:} What are AI software developers' perspectives on redundant code, including how they address it in their current practices and the challenges they encounter?
    
    \textbf{RQ4:} How effective are LLMs in identifying redundant code in AI systems and providing effective optimization techniques?

\section{Expected outcomes}
This doctoral research aims to enhance the understanding of redundant code and propose strategies to optimize them effectively. The expected outcomes include - 
\begin{enumerate}
    \item Understanding the prevalence of redundant code in source code, its impact on overall code quality, and the reasons behind its introduction is essential for improving software development practices. Identifying the factors that lead developers to introduce redundancy—such as time constraints, lack of awareness, or insufficient adherence to software engineering principles—can provide critical insights into addressing the root causes. 
    \item Developing a tool powered by LLMs to eliminate redundant code while preserving its original functionality.
\end{enumerate}

\section{Expected contribution}
To achieve our research goal, we define the following contributions:

\textbf{Analysis of redundant code prevalence: } To analyze the presence of redundant code in source code, we will collect open-source software (OSS) projects focused on AI, following the methodology of Li et al. \cite{li2022exploring}, and leverage three widely used LLMs: GPT-4 \cite{achiam2023gpt}, Gemini \cite{team2023gemini}, and Llama \cite{touvron2023llama}. For each project, we will submit an entire file to the LLM to identify and optimize redundant code. The optimized code is then integrated back into the original codebase, followed by running test cases to ensure all functionalities remain intact. Additionally, we will use static analysis tools to evaluate code quality metrics such as Lines of Code (LOC) \cite{morozoff2009using}, Cyclomatic Complexity (logical complexity) \cite{ebert2016cyclomatic}, and Code Churn (rate of code changes) \cite{shin2010evaluating}. We will document the optimized code that successfully passes the test cases along with its metric values. If the optimized code fails, we will also record the failure details and reasons for the failed test cases.

\textbf{Building a Catalog of Redundant Code Reasons and Identifying Common Patterns: } We will create a comprehensive catalog of reasons behind the introduction of redundant code by analyzing existing studies and gathering insights from developers based on their experience, commit messages, code reviews, and other established resources. Simultaneously, we will leverage LLMs to identify recurring patterns of redundancy in codebases, such as copy-paste effects, repetitive logic, and overused conditional statements. These patterns will be analyzed for their impact on key metrics like the Code Maintainability Index (MI) \cite{welker1997development} and Bug Density \cite{bach2017impact}, providing actionable insights into the most detrimental patterns.

\textbf{Prototype of an automated tool: } We aim to utilize insights from our prior objectives to develop an LLM-agent capable of analyzing a codebase, optimizing files one at a time, reintegrating them into the original codebase, and running test cases to ensure functionalities remain intact. The agent will also analyze failed test cases to identify scenarios where optimizations were ineffective, enabling it to refine its approach for better results. This implementation has the potential to revolutionize how developers identify and optimize redundant code. 

\section{Planned evaluation}
The evaluation of this project involves multiple components. First, we will ensure that the code generated by the LLMs maintains the functional and operational characteristics of the original codebase by executing test cases to verify there are no regressions. A systematic literature review will also be conducted to position our work within the current state-of-the-art and identify relevant research gaps. To understand the reasons behind redundant code, we will collaborate with developers via semi-structured interviews and surveys to validate identified patterns through their feedback and insights. To evaluate the automated tool, we will conduct user studies to gather feedback on its ease of use and effectiveness. Additionally, we plan to use the NASA Task Load Index (NASA-TLX) \cite{hart2006nasa} to assess the cognitive workload associated with using the tool. As the project is in its early stages, the evaluation framework may be refined based on future design decisions.

\section{Limitations}
The proposed research has limitations that may affect its scope and applicability. It focuses on open-source AI projects, limiting generalizability to proprietary systems with different coding practices. The reliance on LLMs like GPT-4, Gemini, and Llama introduces potential biases from training data, potentially overlooking edge cases or uncommon coding patterns. Incomplete test case coverage may fail to detect functionality issues in optimized code. Developer feedback, used to validate redundancy patterns, is subjective and context-dependent. Additionally, reliance on metrics e.g., Code Maintainability Index and Cyclomatic Complexity may not fully capture all aspects of software quality. To mitigate these limitations, in future, the research will incorporate diverse datasets from various domains, prioritize comprehensive test coverage, and refine metrics to include subjective quality factors. Feedback will be gathered from developers with diverse backgrounds to ensure broader validity, and the framework will be iteratively refined based on real-world testing and performance evaluations.

\section{Acknowledgement}
This research is supported in part by the Natural Sciences and Engineering Research Council of Canada (NSERC) Discovery Grants program, the Canada Foundation for Innovation's John R. Evans Leaders Fund (CFI-JELF), and by the industry-stream NSERC CREATE in Software Analytics Research (SOAR).

\bibliographystyle{IEEEtran}
\bibliography{arXiv}

\end{document}